# Adaptively accelerating reactive molecular dynamics using boxed molecular dynamics in energy space

Robin J. Shannon,[1,2] Silvia Amabilino,[2] Mike O'Connor[2,3], Dmitrii V. Shalishilin,[4] David R. Glowacki,[1,2,3]

[1]*Mechanical Engineering, Stanford University, Stanford, CA 94305, USA*
[2]*School of Chemistry, University of Bristol, Bristol, BS8 1TS, UK*
[3]*Department of Computer Science, University of Bristol, BS8 1UB, UK*
[4]*School of Chemistry, University of Leeds, LS2 9JT, UK*

**Abstract**

The problem of observing rare events is pervasive among the molecular dynamics community and an array of different types of methods are commonly used to accelerate these long timescale processes. Typically, rare event acceleration methods require an *a priori* specification of the event to be accelerated. In recent work, we have demonstrated the application of boxed molecular dynamics to energy space, as a way to accelerate rare events in the stochastic chemical master equation. Here we build upon this work and apply the boxed molecular dynamics algorithm to the energy space of a molecule in classical trajectory simulations. Through this new BXD in energy (BXDE) approach we demonstrate that generic rare events (in this case chemical reactions) may be accelerated by multiple orders of magnitude compared to unbiased simulations. Furthermore, we show that the ratios of products formed from the BXDE simulations are similar to those formed in unbiased simulations at the same temperature.

# 1. Introduction

Molecular dynamics (MD) simulations have become a widely used computational tool for understanding molecular problems in a wide range of fields, spanning biochemistry[1], surface chemistry[2], combustion, etc.[3] Such simulations can be used to obtain detailed microscopic insights into molecular behaviour, but the simulation of many real phenomena remains both technically and computationally challenging. The increasing efficiency of force evaluation routines (using molecular mechanics, quantum mechanical approaches, and machine learning) has to some extent served to emphasize another significant difficulty in using MD to simulate molecular change, which arises from the so-called "rare event" problem. Because chemical and molecular change generally occurs on timescales (ns, μs or even ms) which are many orders of magnitude larger than the fundamental timescale of the simulation (fs),[4-10] it is often the case that a significant amount of computational resource is required to observe chemical change within a molecular dynamics simulation.

Over the past few years, a range of rare event acceleration techniques have been described in the literature. Popular approaches include milestoning,[11-12] forward flux sampling,[13-14] transition interface sampling,[15] nested sampling approaches,[16-17] nonequilibrium umbrella sampling,[18] metadynamics [19-20] and boxed molecular dynamics (BXD).[4-10, 21-24] These methods typically require that one has some prior insight into the appropriate collective variable(s) required to accelerate the "event" of interest. For a wide range of chemical problems (e.g., free energy sampling in enzyme catalysed reactions, passage of ions through channels, solution phase chemical reactions, etc.), this is a sensible approach, owing to the fact that one often has a reasonable hypothesis as to which collective variables are likely to play an important role. It's worth pointing out that there is a tradeoff between a given rare event method's efficiency, and the quantity of 'prior' information which it requires. The more information one has to constrain the search space, the more efficient the method; the less information one has to constrain the search space, the more expensive the method. Therefore, in cases where one has reasonable starting hypotheses for how a particular event might happen, the methods outlined above save on valuable computational clock cycles.

For cases where one wishes to accelerate the "discovery" of chemical reactions or molecular change, with limited prior insight and a lack of reasonable starting hypotheses, it is often unclear which collective variables to bias in the first place. In such cases, the challenge is to utilize algorithms in order to generate mechanistic hypotheses. Automated generation of reaction networks is a burgeoning field of research,[25-32] with a range of approaches. For example, some groups utilise MD simulations in order to determine reactive pathways for a given molecular species. [28-29, 31, 33-34] For such approaches, the utility of MD depends on minimising the number of simulation steps and thus the CPU time required to see a reaction event. There have been a number of studies designed to tackle such a problem. For example, metadynamics in conjunction with a SPRINT coordinate representation of a chemical system has been used successfully in this context to explore the combustion mechanism of methanol.[34] There are related general acceleration techniques based on the application of bias potentials such as the hyperdynamics,[35] "accelerated molecular dynamics",[36] and "temperature accelerated dynamics"[37] techniques. There is a separate group of related general acceleration methods based upon parallel tempering or replica exchange.[38-39] Other approaches have used a "piston" style constraint in order to smash species together at high collision energies,[29] while others have simply focused upon high energy MD simulations[28] in order to discover new chemistries.

In this paper we present an alternative method in which we extend the BXD algorithm to adaptively bias the potential energy of the system. Using such an approach, we demonstrate that the 'discovery' of reactive events in MD simulations may be accelerated by several orders of magnitude. A related method called "boxed molecular kinetics" (BXK)[40] has recently been applied to accelerate rare events in kinetic Monte Carlo master equation simulations. This method described herein is essentially the analogue of BXK for the MD domain. This paper is organised as follows. Section 2 focuses on the background and development of the BXD in energy (BXDE) methodology and discusses some practical implementation issues. In Section 3, we utilize this new BXDE method in conjunction with reactive molecular dynamics simulations of an isoprene peroxy system. This is a species of significant atmospheric interest which is known to undergo a number of different reaction pathways[41-43] and as such provides a good test system for comparing reactive events observed from MD simulations with and without the BXDE method. In section 4 we build on previous work[23] and describe methodology for simultaneously implementing both BXD and BXDE constraints. To demonstrate this approach, we consider the aforementioned isoprene peroxy species and demonstrate how addition of an additional BXD constraint limits the dominance of entropically favoured reaction pathways, and thereby encouraging the system to sample a wider range of reaction channels. In Section 5, we present conclusions.

## 2. The BXDE Method

### 2a. Boxed Molecular Dynamics (BXD)

The BXD method[22, 44] has been refined over the last several years and applied[45] to a wide range of systems. In essence, this method works by introducing one or more reflective barriers in the phase space of some MD trajectory, along a particular coordinate or collective variable $\rho(r)$ which is typically a function of the atomic positions $r$ (although it is also possible to formulate $\rho$ so that it is a function of any phase space descriptor, including momenta). These boundaries are used to confine the trajectory to particular regions of phase space, and ultimately to nudge the dynamics along the generalised coordinate into regions of the potential which would be rarely sampled in an unbiased trajectory. Typically the BXD reflective barriers are placed so as to direct the trajectory along some collective variable, which enables sampling of reactant, product, and transition regions of the configuration space. This BXD procedure has recently been generalised to enable adaptive sampling of combinations of collective variables of arbitrary dimensionality;[23] however (for the sake of simplicity) the discussion and figures herein assume that $\rho$ is one dimensional. A schematic of the BXD procedure in 1D is shown in Figure 1. The BXD methodology provides 'box-to-box' rate coefficients. By converging these rate coefficients, the free energy profile along $\rho(r)$ may be obtained. Thus BXD not only accelerates rare events; it also provides a route to recovering both thermodynamic and kinetic information for a particular process.

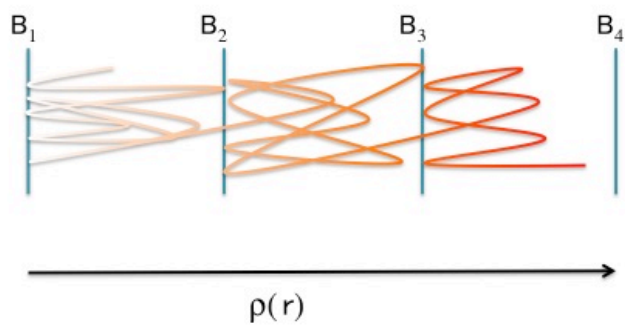

Figure 1: Schematic of the BXD process showing a fictitious trajectory (white indicates short times, orange intermediate times, and red long times) subject to multiple BXD constraints (vertical lines). The constraints are effectively reflective barriers which operate to nudge the trajectory along some generalised co-ordinate $\rho(r)$. After a specified number of hits at a trajectory surface, the constraint/ boundary on the right allows the trajectory to progress and gradually be guided along $\rho(r)$

*2b. BXD in potential energy space*

In this article, we are not concerned with using BXD to obtain ensemble averaged thermodynamics and kinetics, but rather to accelerate the rate at which we observe chemical reactions. Conceptually, it is easy to understand how BXD accelerates rare events. By not allowing a trajectory to sample some specified "box", we effectively place constraints on the phase space volume which it is allowed to sample – i.e., with certain regions of phase space 'off-limits', the trajectory can be forced to sample regions of phase-space which it would otherwise only rarely sample. Work by some of the present authors[23] recently demonstrated that this procedure could be (i) generalised to multidimensional collective variables, and (ii) executed adaptively, in order to make 'on-the-fly' decisions as to where BXD constraints should be located. Here we briefly summarise the foundations of the BXD method, as generalised previously,[23] for the reader's convenience and in order to introduce notation. From an implementation point of view there are two main aspects to the above procedure: (1) How is the trajectory confined within a box? (2) Where should the reflective barriers be placed?

Trajectories are confined within a given box using a velocity inversion procedure when one of the reflective barriers is "hit". We introduce the co-ordinate $\phi(r) = \rho(r) \cdot \hat{n}_j + D_j$ where $\hat{n}_j$ is a unit norm to boundary $B_j$ and $D_j$ is the distance of $B_j$ from the origin. Thus, $\vec{\phi}(r)$ gives a measure of the distance of the trajectory, at any given timestep, from BXD boundary $B_j$. A boundary hit is then defined as a change in sign of $\vec{\phi}(r)$ between $\vec{r}(t)$ and $\vec{r}(t + \Delta t)$ and the BXD constraint can be expressed as:

$$\vec{\phi}(r) \geq 0 \qquad \text{Eq (1)}$$

The velocity inversion procedure works as follows. If at time $t + \Delta t$ the system evolves such that the coordinates pass some reflective BXD boundary $B$, then the system is reset to their values at time $t$ and the velocities are modified to give::

$$\vec{v}'(t) = \vec{v}(t) + \lambda M^{-1} \nabla \vec{\phi}^T \qquad \text{Eq(2)}$$

where

$$\lambda = \frac{-2 \nabla \vec{\phi} \cdot \vec{v}(t)}{\nabla \vec{\phi} M^{-1} \nabla \vec{\phi}^T} \qquad \text{Eq(3)}$$

where $v'(t)$ are the modified velocities which will satisfy Eq. 1 and $M$ is $3N$ by $3N$ matrix of atomic masses with $N$ being the number of atoms in the system. These equations (Eq2,3) simply revert the component of the velocity in the direction of reaction coordinate.

To accelerate general reactive events, there are two criteria which much must be satisfied: (1) the molecule must be in the correct configuration (i.e., the entropic penalty to reaction must be overcome); and (2) the potential energy of the species (located in the correct subset of modes) must exceed the energy barrier to reaction (i.e., the enthalpic penalty to reaction must be overcome). To date, BXD applications have mostly taken a sort of 'entropic' approach to accelerating rare events: it constrains a molecular system in a region of configuration space which is close to some transition region until spontaneous energy fluctuations enable the system to cross over into the next 'box'. An alternative – and the one which we develop herein – is effectively an enthalpic approach: by confining a molecular system to regions of high potential energy, we increase the probability of observing spontaneous energy fluctuations which enable a molecule to reach the correct configuration which promotes reaction. To this end we have introduced a BXD bias along the potential energy (E) of the system, which we hereafter refer to as 'BXDE'. By 'scanning' through potential energy 'boxes', we can identify the energetic 'windows' at which different chemical reactions channels switch on or off.

For BXDE, the velocity inversion is trivial, owing to the fact that $\nabla \phi$ is simply given by the forces used for propagating the MD. Figure 2 (and its corresponding supplementary video) shows the BXDE procedure in the case of a simulation of a point particle on the so-called the 'Müller-Brown potential'.[46] The different coloured zones are simultaneously energy contours and BXD boundaries. Purple corresponds to low energies, and red to high energies. Starting from panel A, at the beginning of the simulation the system is in one potential energy well. It explores that region until it has hit the boundary a user-specified number of times. At that point, the system is allowed to proceed to the next region as shown panel B. 'Off-limit' regions which have already been explored are coloured in grey. This procedure continues in panels C and D.

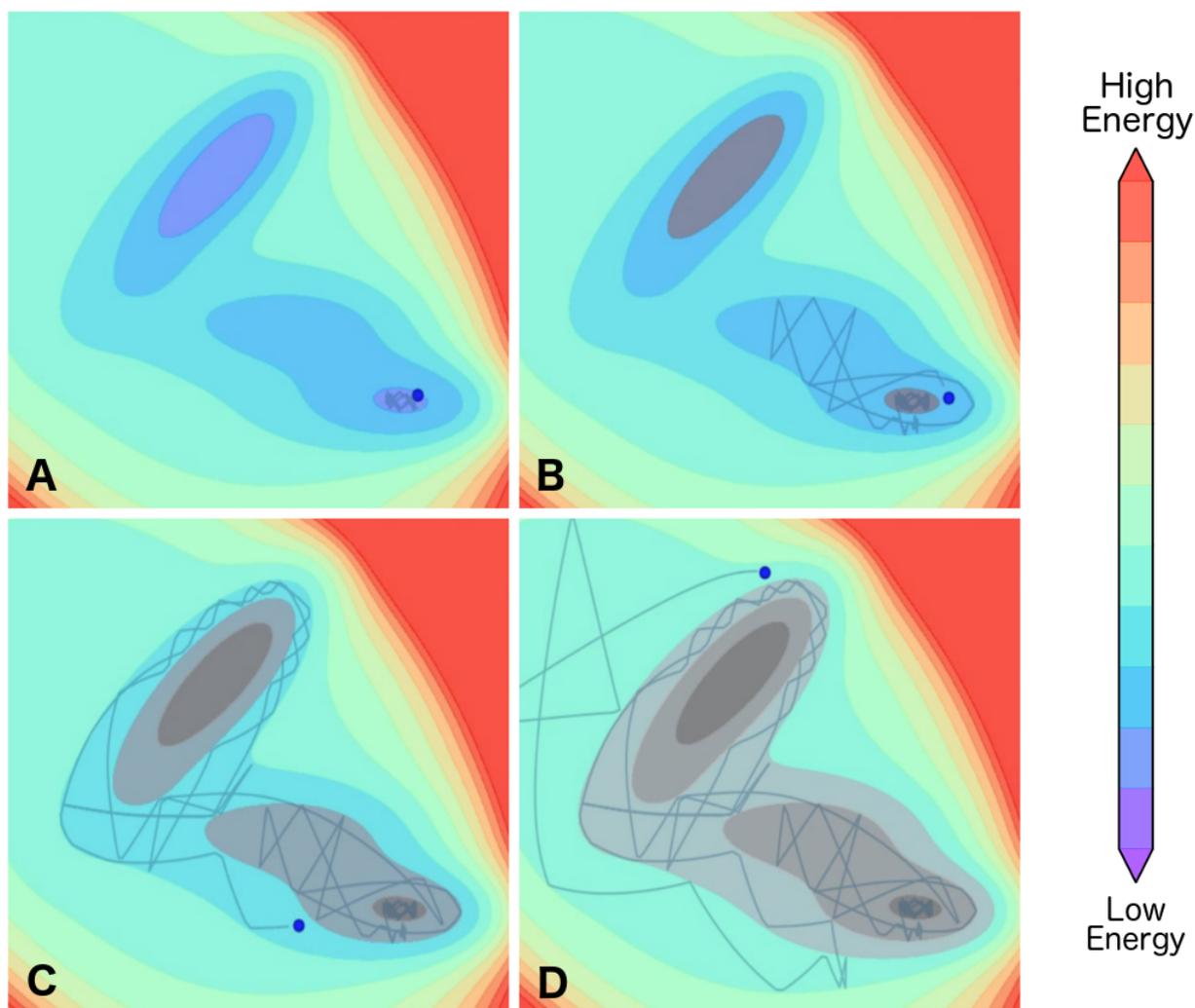

Figure 2: Schematic of the BXDE process from trajectories of a fictitious point particle on the Mueller Brown potential energy surface utilising an NVE approach. Purple means low energy and red high energy. The particle starts in one

potential energy well and explores one region at a time. The different coloured regions are separated by BXD boundaries. The areas that can no longer be explored are coloured in grey. The trajectory time series progresses from A to D.

## 2c. BXDE implementation issues

Fig 2 illustrates BXDE run in an NVE ensemble; however it is also possible to run in an NVT ensemble. In general, we have found the NVT ensemble of BXDE to be more robust for two reasons. First, the system can sometimes get trapped in the vicinity of a BXDE constraint. Such a scenario is shown in Fig 3, where the system (shown as a black dot) has a velocity vector (black line) which would take it beyond a BXDE constraint, and therefore requires velocity inversion (grey line); however, the velocity inversion procedure also causes the system to cross a BXDE constraint. This leads to an infinite loop of inversions (back and forth), trapping system near the boundary. Systems with few degrees of freedom (such as a particle on a 2d Mueller-Brown potential) are particularly susceptible to this problem. Second, running BXDE in the NVE ensemble leads to a scenario where – as the system progresses through regions with higher potential energy – the kinetic energy necessarily decreases (because the total energy must be conserved), and the system explores higher energy regions significantly more slowly.

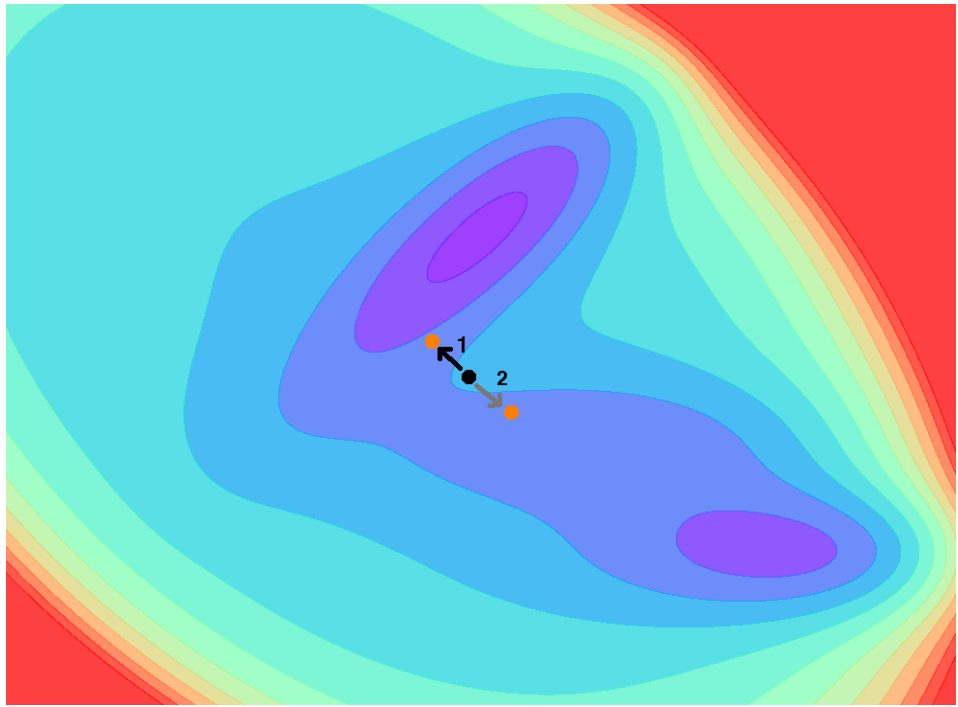

Figure 3 - Diagram showing a model 2D system where a particle approaches a boundary (velocity vector 1 black line) and after the inversion procedure it takes another step outside the boundary (velocity vector 2 grey line). This results in an infinite loop of inversions.

In the NVT ensemble (using e.g., a Langevin thermostat), these problems do not arise. Figure 4 (and its corresponding supplementary video) illustrates the impact of adding a Langevin thermostat to the simulation of the model system in a Müller-Brown potential. In the Langevin thermostat, temperature is conserved through attempting to mimic dynamical effects of a heat bath – i.e., viscous drag and random collisions with bath particles. The corresponding equation of motion is as follows:

$$m\,a(t) = F(t) - m\,\gamma\,v(t) + \chi(t) \qquad \text{Eq. 4}$$

where $m$ is the mass of a particle, $a(t)$ is the acceleration, $F(t)$ is the force acting on the particle, $\gamma$ is a friction coefficient, $v(t)$ is the velocity of the particle and $\chi(t)$ is a random force.[47] The random forces included within the Langevin bath significantly reduce the likelihood of the system being trapped near a boundary, with random collisions ensuring that the system is 'kicked' out of any traps.

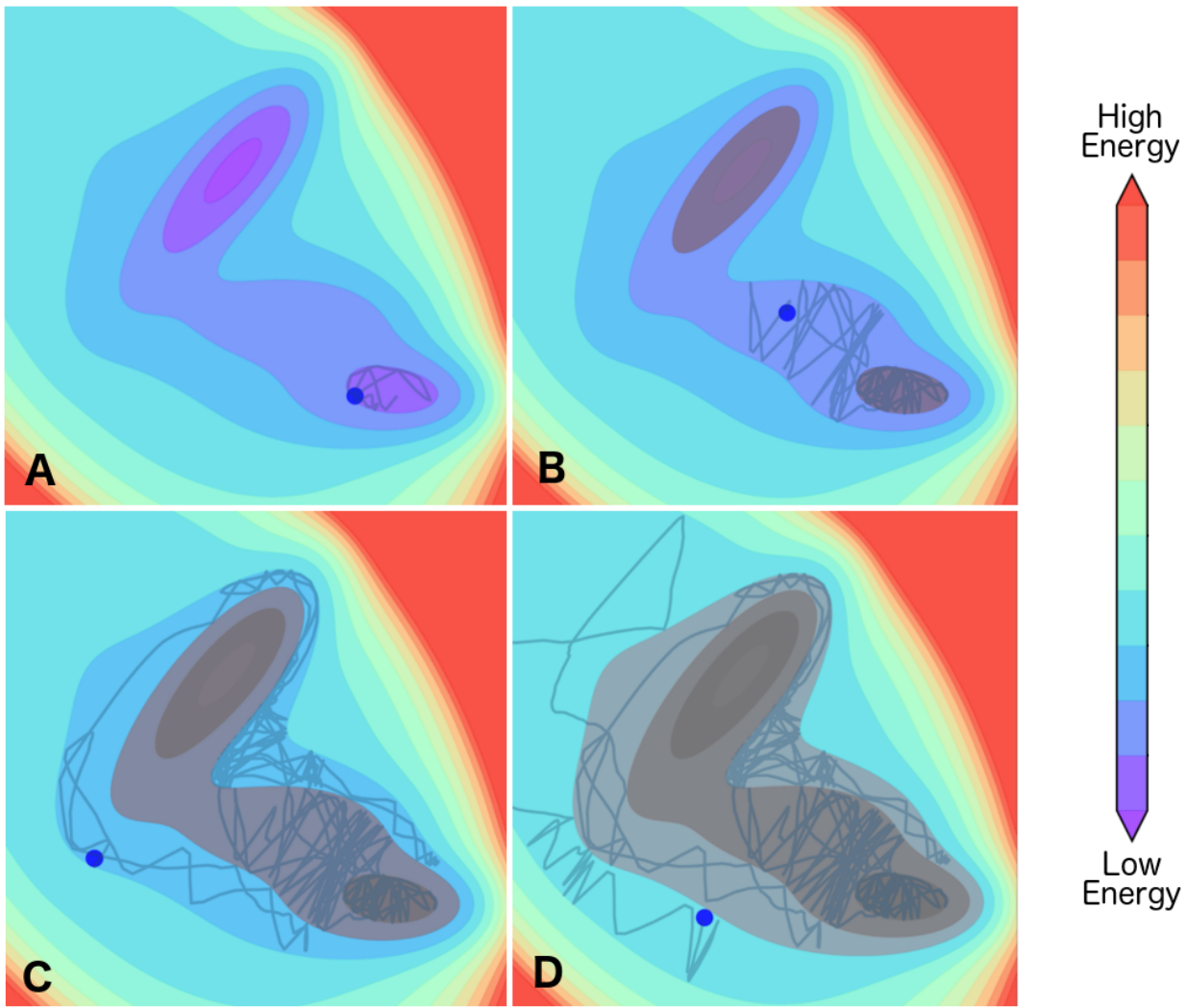

Figure 4: Simulation of a particle in a Müller-Brown potential. In this case the trajectories are run in the NVT ensemble with a Langevin thermostat.

## 2d. Adaptive Placement of BXDE Boundaries

One of the most important considerations when implementing any BXD procedure concerns where to place the boundaries. When studying a system for the first time, it often requires some exploratory 'trial-and-error' runs in order to determine the optimal placement of BXD boundaries. To eliminate this initial 'trial-and-error' aspect of BXD, O'Connor et al.[23] recently described an 'on-the-fly' adaptive scheme for boundary placement along specified configuration space collective variables. In BXDE, the adaptive boundary placement scheme is illustrated in Figure 5. The molecular dynamics simulation is allowed to run freely, keeping track of $V(t)$ at each timestep. When $V(t) > V(t - dt)$, the maximum value of the potential energy ($V_{max}$) is updated – i.e., $V_{max} \leftarrow V(t)$. After a user defined number of steps $i_{samp}$ have elapsed, a BXDE boundary is set at the current value of $V_{max}$ – i.e., $V_{BXDE} \leftarrow V_{max}$. During the adaptive procedure, $V_{BXDE}$ bounds from below only, such that the trajectory is allowed to cross the boundary at the next timestep where $V(t) > V_{BXDE}$. At subsequent timesteps, the system is confined to potential energies $V(t)$ greater than $V_{BXDE}$ by invoking the BXDE velocity inversion procedure described above. This procedure enables the system potential energy to steadily increase.

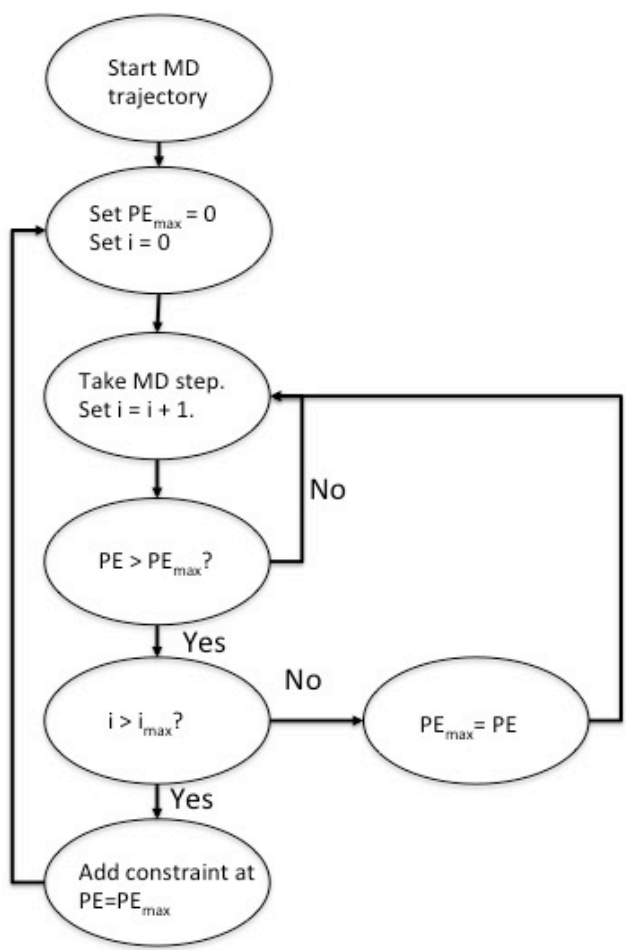

Figure 5: Flowchart describing the adaptive placement of constraints in the BXDE procedure.

While not considered in the current work, the BXDE methodology can also be used to place reflective boundaries that bound from above – i.e., to reflect the trajectory towards regions of lower potential energy. This could be used in the above adaptive scheme to define some $V_{upper,}$ which is never exceeded. This type of approach combined with an NVE ensemble could be used to study chemical reaction in a well defined window of energies. Conceptually, such an approach would map well onto micro-canonical transition state theory calculations in particular the aforementioned BXK methodology[40].

## 3. Reactivity in an Isoprene Peroxy System
*3a. Simulation Details*

To test the BXDE approach, we have implemented it within a locally modified version of the DFTB+ code[48], enabling us to investigate reactive MD trajectories of the isoprene peroxy radical (OHCH$_2$C(CH$_3$)=CHCH$_2$OO) whose structure is shown in scheme 1. A repository containing the modified DFTB+ code may be found at https://github.com/RobinShannon/BXDE_DFTB. This species is a key intermediate in the isoprene atmospheric oxidation sequence, and therefore plays an important role in atmospheric chemistry, owing to the fact that isoprene is amongst the most abundant volatile organic hydrocarbons (VOCs) in the troposphere [49]. As a test for BXDE, isoprene peroxy offers an interesting system, owing to the fact that it is known to have a number of reactive isomerisation and dissociation pathways[42]. To initialize our BXDE runs, the geometry of the isoprene-peroxy radical was optimised using DFTB with an SCC correction and the mio-1-1 parameter files.[50] MD trajectories were then performed with a 0.1 fs timestep using an NVT ensemble and a Langevin thermostat with a friction coefficient of 0.05 ps$^{-1}$.

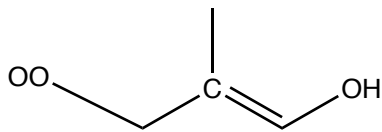

Scheme 1: Structure of the isoprene peroxy radical.

An adaptive approach was used for the placement BXDE boundaries as detailed in Figure 5. Simulations were performed with $i_{samp}$ values of both 100 and 1000, in order to evaluate the extent to which this parameter impacted the observed product channels. Simulations were performed until a reaction was observed as defined in section 3b, at which point the BXDE boundaries were removed to allow the dynamics to evolve naturally in the product region. A further 1000 MD steps were then performed in order to confirm the product species identity. This product geometry was then optimised at the dftb-scc level of theory and the canonical SMILES string for this geometry was recorded for bookkeeping purposes (computed by the openBabel [51] software).

*3b. Reaction Criteria*

To keep track of when a chemical reaction occurred, we identified bond making and breaking events using the approach recently described by Martinez Nunez.[28] In this procedure two matrices are defined, $d$ and $d^{REF}$. The first matrix has elements $d_{ij}$ equal to the distance between atoms $i$ and $j$ in the system; the second matrix has elements $d_{ij}^{REF}$, consisting of pre-defined ideal bond distances between atoms $i$ and $j$. In this work the ideal bond distances were defined according the atoms involved and were set to the following values in Table 1.

| $ij$ | $d_{ij}^{REF}$ / Angstrom |
|---|---|
| CC | 1.6 |
| CH | 1.2 |
| CO | 1.6 |
| OO | 1.6 |
| OH | 1.2 |
| HH | 0.8 |

Table 1: Ideal bond distances which make up the elements of the $d^{REF}$ matrix

Using these two matrices, we form a connectivity matrix with elements $C_{ij}$, where:

$$C_{ij} = \begin{cases} 1 \text{ if } d_{ij} < d_{ij}^{REF} \\ 0 \text{ otherwise} \end{cases} \qquad \text{Eq. 5}$$

For the starting structure, this matrix identifies whether two atoms are bonded ($C_{ij} = 1$) or non-bonded ($C_{ij} = 0$). At each time step, the current bonding structure (given by $d$ and $d^{REF}$) of the system is compared with the reactant bonding structure (given by $C$) to monitor for reaction. Specifically, a reaction is then considered to occur if for an atom $i$:

$$\max(\delta_{in}) > \min(\delta_{ik}) \; ; \; \delta_{ij} = \frac{d_{ij}}{d_{ij}^{REF}} \qquad \text{Eq. 6}$$

Here index *n* runs over atoms bonded to *i* ($C_{ij}$ matrix elements equal to one) and index *k* runs over atoms which do not have a bond to *i* ($C_{ij}$ matrix elements equal to zero). In this work we introduce the additional constraint that the criteria above be met consistently for 50 MD timesteps, to filter out extremely short-lived transient reaction events, and ensure that a bond breaking / bond forming process had indeed occurred. There is no need to recalculate the elements of *C* at the product geometry, since in the current work simulations are stopped once a reaction product has been identified.

*3c. Results*

Simulations were performed with and without the BXDE procedure at a number of temperatures between 500 K and 4500 K. Figure 6, which indicates the temperature variation in the 500K BXDE simulations, shows that the system is close to the target temperature of 500K.

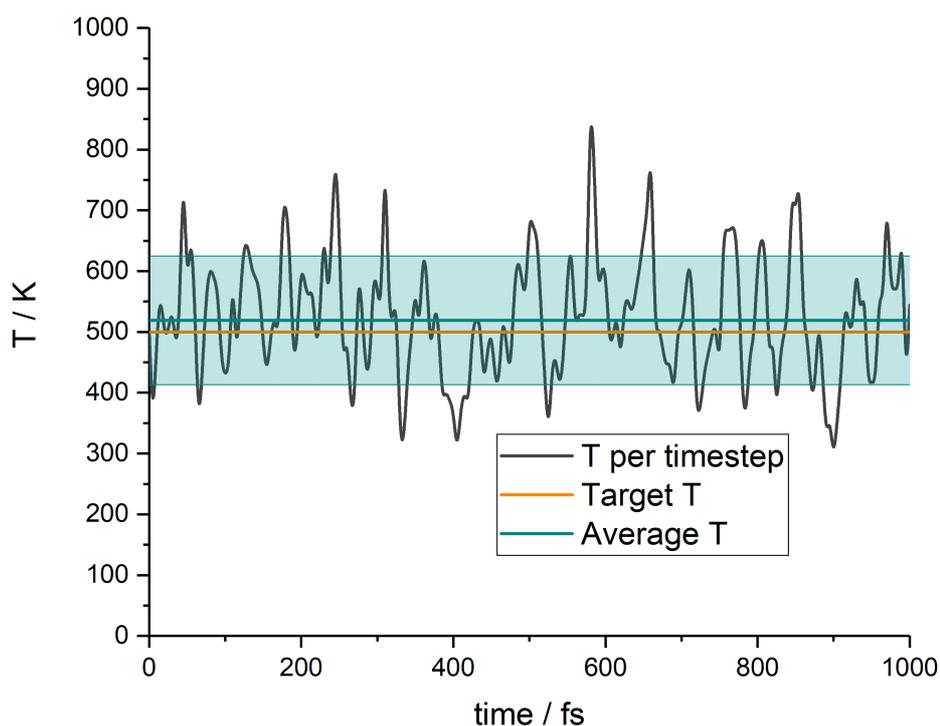

Figure 6: Temperature as a function MD simulation time in fs, for BXDE simulations at 500K. Here an NVT ensemble has been used with a temperature of 500 K and a friction coefficient of 0.05 ps$^{-1}$. The error bars shown correspond to 2σ standard deviations.

Figure 7 shows the average time elapsed, prior to observation of a reactive event (as defined by the criterion in Eq. 4 and Eq. 5), obtained (where possible) from 50 different reactive trajectories at temperatures of 500 K, 1500 K, 2500 K, 3500 K and 4500 K. Figure 7 shows the degree of acceleration in reactive events which can be achieved by utilising BXDE. As would be expected, reactive events are observed more rapidly with an $i_{samp}$ value of 100 though this difference is only significant at the lowest temperature. Since we were unable to observe any reactions at 500 K without BXDE, we estimated the non-BXDE 500K value from microcanonical transition state theory calculations using the master equation code MESMER.[52] The input file for these calculations is given in the supplementary information, and these calculations yield a rate coefficient of $1.29 \times 10^5$ s$^{-1}$ – i.e., the MD simulations would need to run for an average of $7.75 \times 10^9$ fs in order to observe reaction. Error bars were calculated using Poisson statistics, with the sample standard deviation calculated as $\sigma = \frac{s}{\sqrt{n}}$ where s is the variance in the sample data and n is the sample size, in this case 50. Errors shown are 2σ.

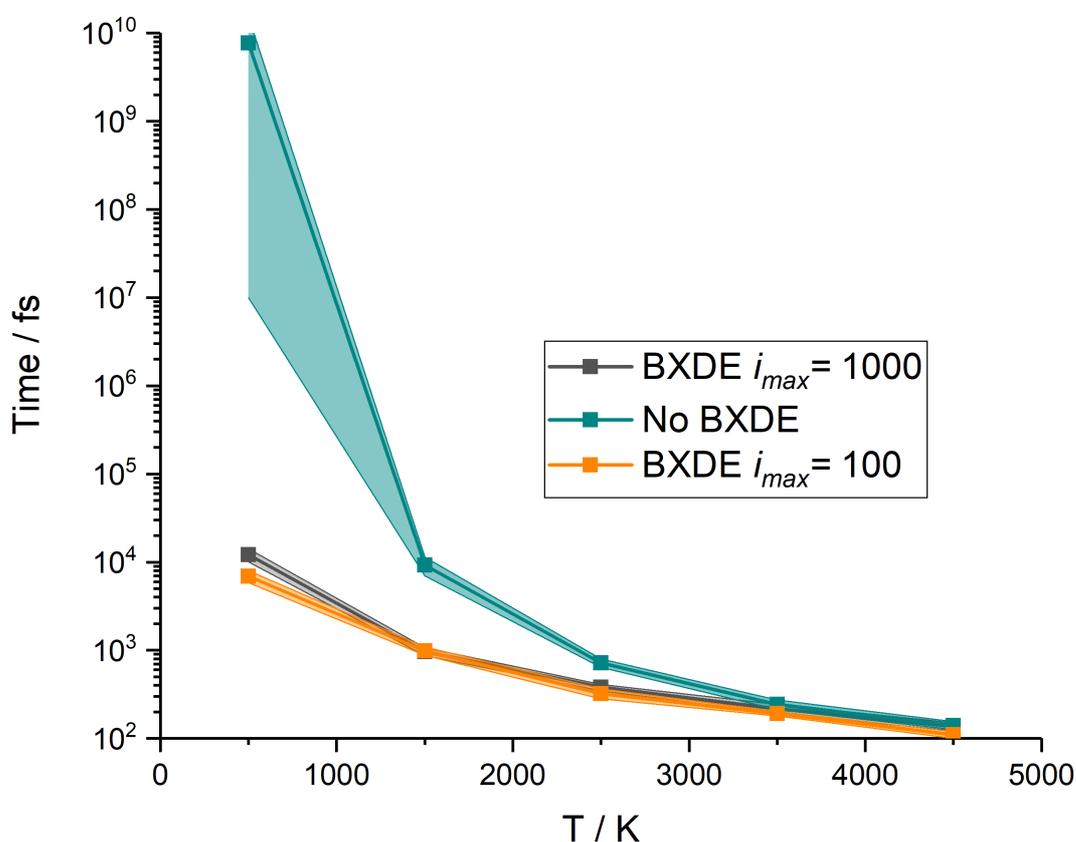

Figure 7: Average time in fs before reaction is observed at different temperatures.

Broadly speaking, Fig 7 lends credibility to a relatively common approach for accelerating chemical reactions – i.e., simply running dynamics at extremely high temperatures or energies.[28-29] However there are two interrelated problems associated with using high energy trajectories: (1) Because the object of a chemical dynamics study is usually to understand reactivity at a particular energy or temperature range, extremely high temperature dynamics – which leads to an explosion in the number of possible reactive pathways – are unlikely to be representative of those pathways which are accessible under typical experimental conditions; and (2) Extremely high temperatures and energies tend to favour pathways with significant entropic benefits over those with low enthalpic barriers – i.e., loose transition states corresponding to dissociation pathways are observed far more frequently than lower energy isomerisation pathways with tighter transition states.

Figures 8 compares the different product channels observed from BXDE simulations at 500K and 4500K, illustrating the point discussed above. The minor channels, not shown in figure 8 are listed in the online supporting information. The chart in the panel A shows the observed 500K products. At 500 K, the loss of $O_2$ is the dominant channel. There is also a small $HO_2$ loss channel though this process involves isomerisation prior to dissociation (hence the I/D label); all the other channels observed correspond to unimolecular reactions involving the transfer of a hydrogen or oxygen atom from one part of the peroxy to another. Previous high level studies of this isoprene peroxy system have shown that hydrogen transfer from C to O (giving I1) has the lowest barrier to reaction.[42] It occurs as the second most frequent product channel in the 500 K BXDE simulations. For the 4500 K case (shown in panel B) a total of 18 product channels were observed and only three of these overlap with the products observed from simulations at 500K. The corresponding chart splits these channels according to the number of molecular fragments formed. Unimolecular reactions typically have the lowest energy barriers but they are entropically unfavourable owing to the fact that they are relatively 'tighter', and therefore are not observed during the 4500 K BXDE runs. All of the observed product channels instead involve dissociation of the isoprene peroxy into multiple (anywhere from 2 – 4) fragments. The hydrogen transfer channel in Figure 8a (I1), which is known to be the lowest energy channel, is in fact not observed at 4500K.

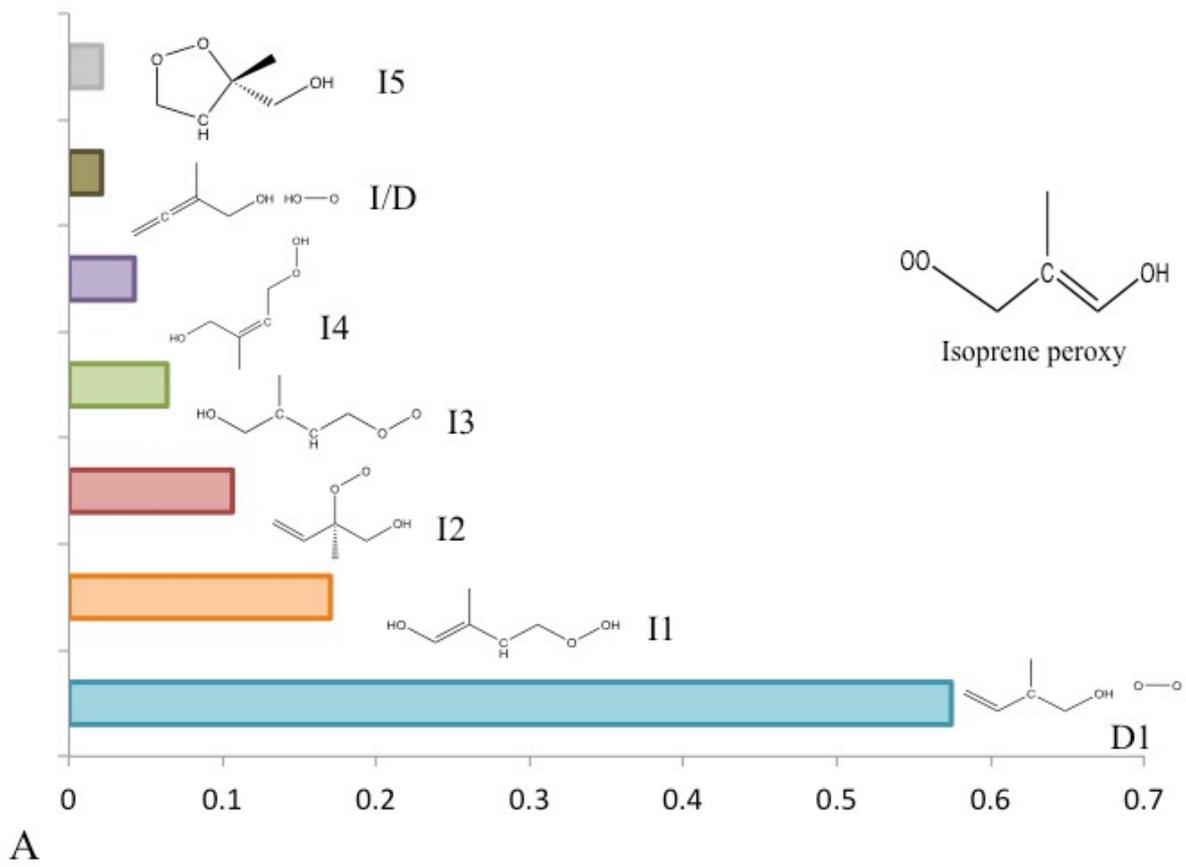

A

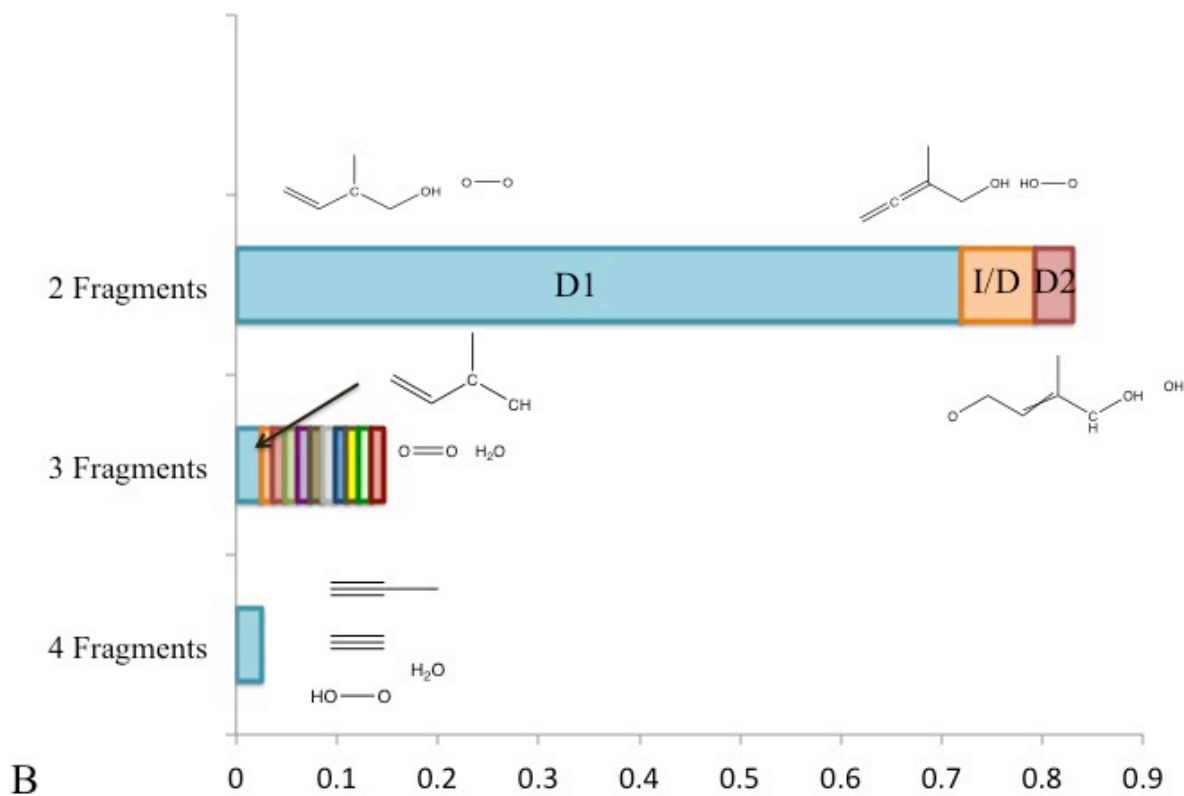

Figure 8: Breakdown of product channels from (A) 500K and (B) 4500K. In the 500K case, each bar shows a separate product channel, whereas in the 4500K case, the bar chart instead distinguishes the product channels by the number of fragments formed with each bar broken down into individual products. Product structures are shown for all channels with more than one occurrence and those channels found in (B) which are also found in (A) are labelled. Minor product channels from the 4500 K simulations can be found in the online supporting information

The two major reaction products shown in Fig 8A (H transfer I1 and $O_2$ loss D1) are shown in Figure 9 alongside two other low energy channels: loss of OH to (D2) and H transfer from O to O to form (I2). To visualise what is happening in the BXDE trajectories, Figure 9 shows the potential energies sampled during a 2500 K BXD run, along with adaptively placed BXDE boundaries. Also shown are the reaction thresholds for the 4 reactions mentioned above. It should be noted that optimisation of the saddle points corresponding to the H transfer processes on the left hand side was not possible using the DFTB method as implemented in the G09 suite of *ab initio* programs,[53] perhaps due in part to the lack of analytical second derivatives for the DFTB method. As such the barrier heights are estimated from single point DFTB calculations at the optimised geometry of higher level M06-2x /6-311+G(3d,2p) transition states.

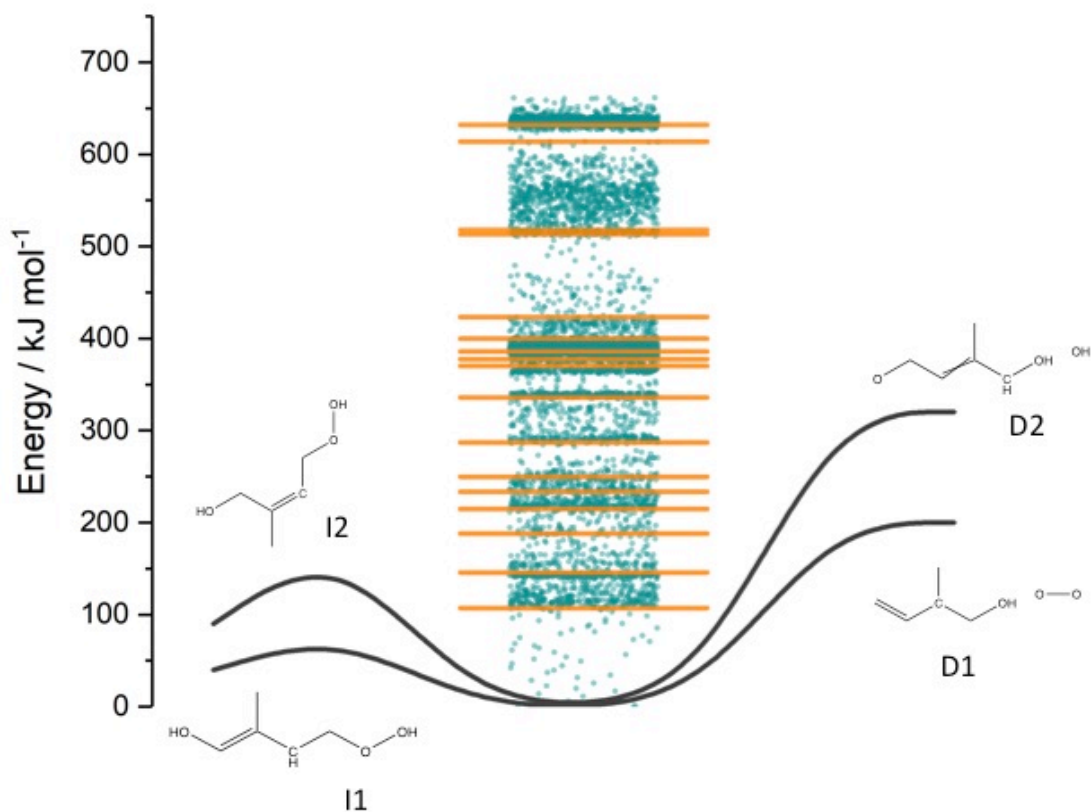

Figure 9: Data schematic from a 2500K BXDE simulation. Also shown are estimated barrier heights for the four lowest energy product channels as calculated at the DFTB level of theory. The channels to the left correspond to hydrogen transfer unimolecular channels and the channels on the right are dissociations lacking defined saddle points.

*3d. Sampling enthalpic vs entropic channels*

In what follows, we show that a simple statistical rate theory analysis is able to explain the dominance of entropic channels observed in BXDE runs at high energies. Using the estimated barrier heights shown in Fig 9, we performed microcanonical statistical rate theory calculations using the master equation solver MESMER,[52] a more detailed description of which can be found elsewhere.[54-55] In our master equation calculations we compare the hydrogen transfer channels on the left of Figure 9 and the dissociation channels on the right side of Figure 9. The MESMER input used to carry out these calculations is given in the supplementary information.

Figure 10 shows the energy resolved rate coefficients ($k(E)s$) for the product channels in Fig 9 calculated using MESMER. The plot clearly shows that the lowest hydrogen transfer isomerization channel dominates at low energies, but that both dissociation channels dominate at higher energies (by 3 orders of magnitude in the case of D1). This plot also demonstrates why the BXDE trajectories need to be so far in excess of the lower energy thresholds before observation of reactive events: in order for a reaction to be observed on the order of 1000 fs, a rate coefficient on the order of $10^{12}$ s$^{-1}$ is required. This region is only reached at energies more than 100 kJ mol$^{-1}$ above the O$_2$ dissociation energy.

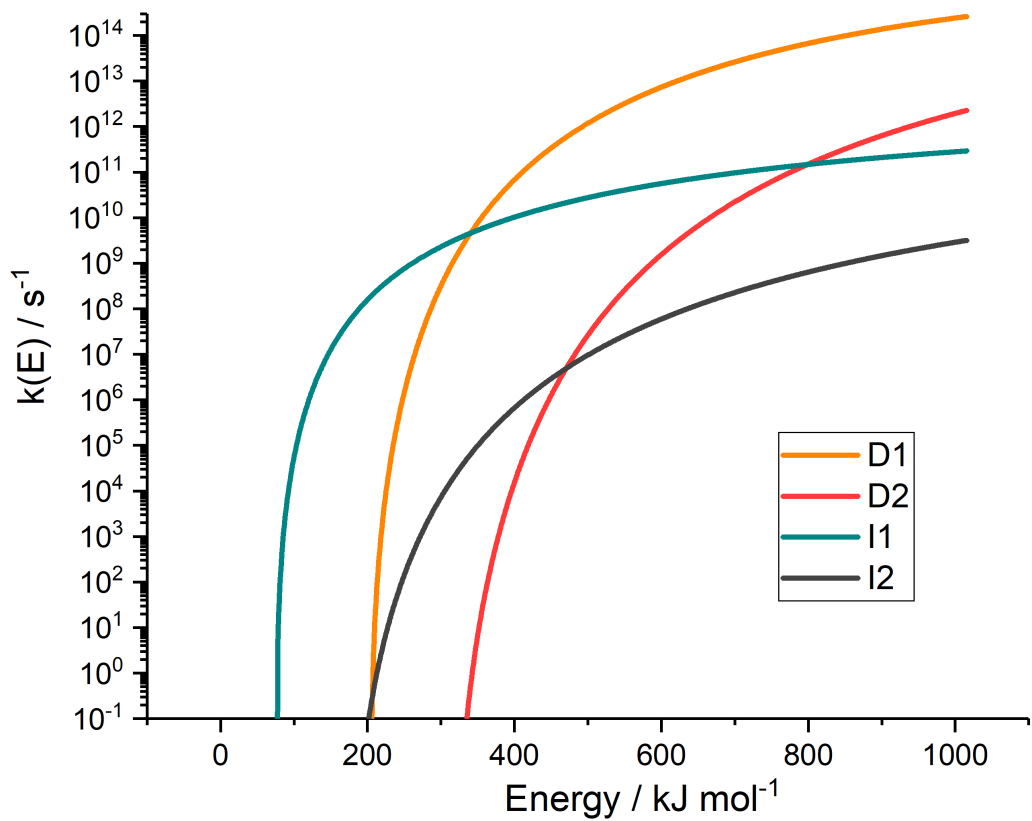

Figure 10: Calculated *k(E)*'s from master equation simulations of the lowest energy reaction paths in the isoprene peroxy system. The labelling follows that show in Figure 9.

To explore the extent to which BXDE simulations are representative of unbiased simulations at the same temperature, we have analysed the product yields of the dominant H transfer and $O_2$ loss channels observed in the MD simulations. These calculations were performed at 500 K, 1500 K, 2500 K, 3500 K and 4500 K. Figure 11 shows the product yields for I1 and D1 as a function of temperature for the unbiased MD simulations and the BXDE simulations with $i_{samp}$ values of 1000 and 100. The error bars on these product yields are taken from a multinomial distribution (as often used when analysing product channels from stochastic trajectories[56]), given by the following expression:

$$\sigma = \sqrt{\frac{1}{N}f(1-f)} \qquad (Eq.8)$$

where N is the sample size (50) and *f* is the fractional yield. The errors bars shown in Fig 11 are 2σ. These calculations demonstrate that product yields are statistically identical between the MD simulation types, and it can be observed that all simulations follow the same temperature dependence. While the error bars are large due to the relatively small sample size, the results suggest that product yields obtained using low temperature BXDE simulations do provide a qualitatively meaningful picture of low temperature reactivity.

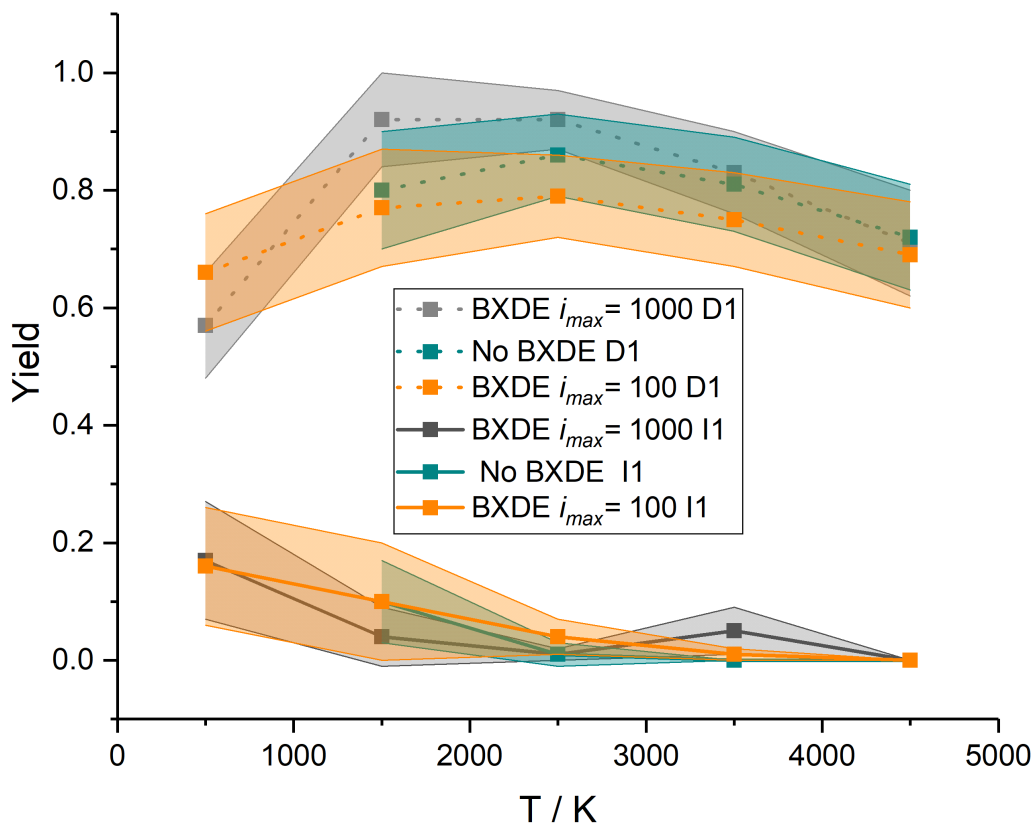

Figure 11: Branching ratios of two important channels ($O_2$ dissociation and H transfer) for the $OHCH_2C(CH_3)=CHCH_2OO$ system. These are shown as a function of temperature and compare non-BXDE simulations with BXDE simulations using $i_{samp}$ values of 1000 and 100.

## 4. Multiple BXD constraints

### *4a. Motivation and Implementation*

The analysis above shows that entropic channels dominate at high energies and temperatures. To avoid this dominance of entropic channels, this section outlines our method for adding additional constraints. Since BXD treats each collective variable individually, additional constraints mean that multiple BXD boundaries need to be enforced (as shown in Figure 12). Enforcing multiple constraints in both energy and collective variable space is distinct from the multidimensional collective variable approach presented in our previous work. In that work, any given hyperplane was defined as a linear combination of collective variables. This ensured that hyperplanes did not intersect within the dynamically sampled space, and hence only one boundary could be crossed at a time. In the present application, it is possible that the boundaries intersect in collective variable space and so a trajectory may cross multiple boundaries in several collective variables in a given time step.

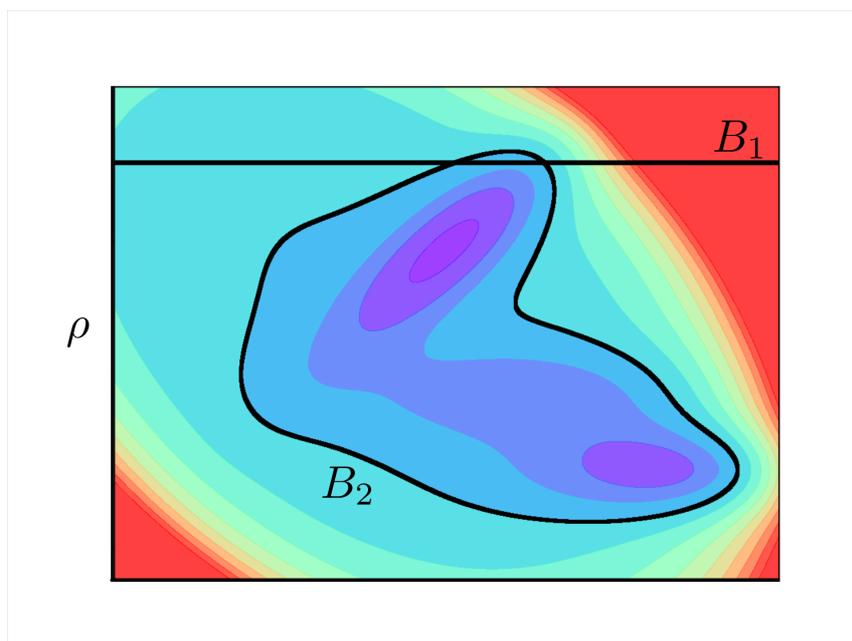

Figure 12: Schematic illustrating the placement of two intersecting BXD boundaries, $B_1$ and $B_2$. $B_1$ restricts the trajectory to particular values of $\rho$ in collective variable space, while $B_2$ restricts the trajectory to a particular energy range. At the points of intersection of the two boundaries, BXD constraints for both must be enforced simultaneously.

For a system of $K$ nonholonomic constraints, we define a quantity associated with each collective variable, as described in section 2a. With these $K$ constraints $\nabla \vec{\phi}_K$ is generalised to a $K$ by $3N$ matrix $\nabla \boldsymbol{\phi}$ and $\lambda$ is a replaced with a vector of size $K$, denoted $\vec{\lambda}$. The modified velocities must now satisfy the following system of linear equations, which we here express in matrix form:

$$\nabla \boldsymbol{\phi}[(t) + \vec{v}'(t)] = 0 \qquad \text{Eq. 9}$$

Substitution of Eq.1 into Eq, 9 gives the matrix equation:

$$\nabla \boldsymbol{\phi}[\vec{v}(t) + \vec{\lambda} \mathbf{M}^{-1} \nabla \boldsymbol{\phi}] + \nabla \boldsymbol{\phi} \vec{v}(t) = 0 \qquad \text{Eq. 10}$$

This defines a system of linear equations, which can be solved to give $\vec{\lambda}$. A worked analytical solution to the current case where $K = 2$ is given in Appendix 1.

### *4b. Isoprene Peroxy Simulations With Multiple BXD Constraints*

Simulations were once again performed for the isoprene peroxy species described in Section 3. These simulations were performed at 500 K with an $i_{samp}$ value of 100 and were carried out in an almost identical manner to those described in Section 3, with the only difference being the addition of a second BXD constraint, constraining the C-OO bond distance of the isoprene peroxy molecule. We placed a reflective BXD boundary at 1.6 Angstroms, to prevent bond cleavage and avoid the dominance of entropically favoured channels. Figure 13 shows a chart of the product channels, which can be compared directly to Figure 8A. One observes that neither the $O_2$ nor $HO_2$ loss channels are present due to the additional BXD constraint. Instead, the two dominant channels are the two H transfer processes, which respectively yield I1 and I2 (the two lowest energy channels on the left side of Figure 9). The more minor unimolecular channels from Figure 8a are no longer observed, although due to the relatively small sample size it is difficult to draw firm conclusions regarding this. We also observe some new channels: namely OH (D2) and O atom loss. These results provide a clear demonstration that the second BXD constraint reduces the number of product channels, avoids the dominance of entropically favoured channels, and thereby enhances the sampling of the lowest energy unimolecular channels.

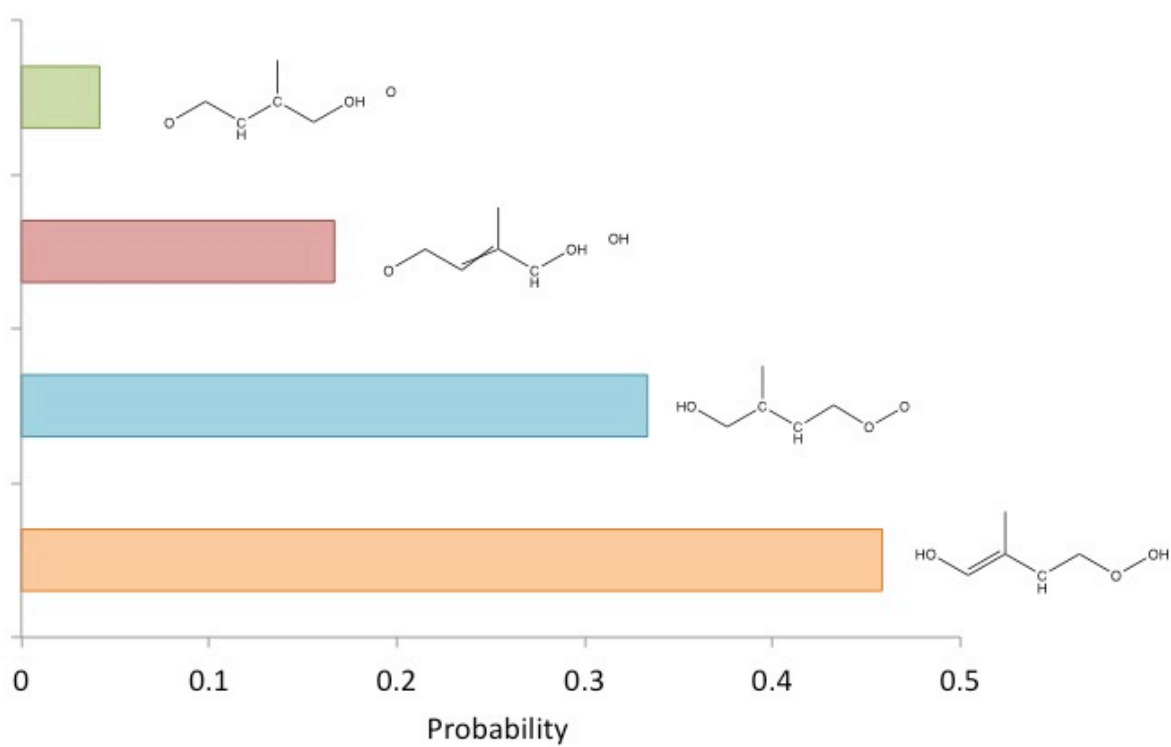

Figure 13: Product channels from 500K BXDE simulations with an additional BXD constraint stopping $O_2$ dissociation. The products formed are labelled with the corresponding structure.

## 5. Conclusions

In this paper we presented a new methodology based upon the BXD procedure, which can accelerate the observation of chemical reactivity in MD simulations by more than 6 orders of magnitude compared to unbiased trajectories. This method allows chemical reactions to be sampled efficiently whilst giving reaction product yields which are consistent with the temperature at which the MD simulations are run. This method allows for rapid sampling of reaction mechanisms via MD and is particularly easy to implement compared to related methodologies.[34-36] This method strongly overlaps with the recently developed boxed molecular kinetics (BXK) [40] approach. Together these methods are ideal for automatically mapping and quantifying the kinetics of complex reactive or conformational networks of the types found in combustion modelling,[3] protein folding[1] and countless other fields.

We have also demonstrated how the BXDE approach can be combined with additional collective variable constraints, in order to avoid the scenario where entropically favoured channels dominate compared to important low energy channels. This multi-constraint approach allows traditional BXD constraints to be coupled with BXDE in order to gain greater control of which reactive events are observed. As demonstrated here, this can be used to remove a particularly dominant reaction path to allow a more detailed investigation of other channels. This approach could also be used for examining the reactivity of weakly bound complexes via MD since a BXD constraint can prevent dissociation of the complex while the BXDE constraint accelerates the observation of chemically reactive processes, which the complex may undergo. In the future the use of multiple constraints could be implemented in an automated manner as different reaction pathways are found. For example, it should be possible to add BXD constraints 'on-the-fly' as reactions are discovered, and thereby encourage the system to sample increasingly minor reaction channels.

In summary the two tools presented here are ideally suited to the rapid exploration of reaction or conformation space within MD whilst maintaining temperature or energy resolution in the observed pathways. The BXDE method is simple and computationally cheap to implement and we demonstrate that it can lead dramatic increases in the rate at which rare events are observed. Work is in preparation to form a new dynamics-based reaction discovery framework utilising the new methodologies described here and there are also plans to apply the BXDE method to conformational / rare event sampling in protein simulations.

## Acknowledgements

DRG acknowledges funding from the Royal Society as a University Research Fellow and also from EPSRC Program Grant EP/P021123/1. Funding for RJS was provided by the US Air Force Office of Scientific Research (AFOSR) under Contract No. FA9550-16-1-0051 and researcher mobility grant from the Royal Society of Chemistry. Funding for SA is from the EPSRC Centre for Doctoral training, Theory and Modelling in Chemical Sciences, under grant EP/L015722/1. MOC is funded jointly by EPSRC and Interactive Scientific Ltd. through an industrial case studentship. This work was carried out using the computational facilities of the Advanced Computing Research Centre, University of Bristol - http://www.bris.ac.uk/acrc/

# Appendix 1

In the two dimensional case the matrix equation represented by Eq. 10 can readily be solved as the system of linear equations:

$$\nabla\vec{\phi}_1 \lambda_1 \mathbf{M}^{-1} \nabla\vec{\phi}_1 + \nabla\vec{\phi}_1 \lambda_2 \mathbf{M}^{-1} \nabla\vec{\phi}_2 = -2\nabla\vec{\phi}_1 \cdot \vec{v}(t)$$
$$\nabla\vec{\phi}_2 \lambda_1 \mathbf{M}^{-1} \nabla\vec{\phi}_1 + \nabla\vec{\phi}_2 \lambda_2 \mathbf{M}^{-1} \nabla\vec{\phi}_2 = -2\nabla\vec{\phi}_2 \cdot \vec{v}(t) \qquad \text{Eq. A1}$$

Here $\nabla\vec{\phi}_1$ is a vector of elements of the first row of $\nabla\boldsymbol{\phi}$ (the BXDE constraint vector) and $\nabla\vec{\phi}_2$ is a vector of elements of the second row of $\nabla\boldsymbol{\phi}$ (the derivative of the centre of mass separation between the isoprene fragment and the $O_2$ group.

Since the coefficients of $\lambda_1$ and $\lambda_2$ are scalars we can simplifiy Eq.11 to:

$$a\lambda_1 + b\lambda_2 = -2c$$
$$d\lambda_1 + e\lambda_2 = -2f \qquad \text{Eq. A2}$$

Then, multiplying the top by $d$ and the bottom by $a$ and then subtracting the bottom from the top we get:

$$\lambda_2 = \frac{-2cd + 2af}{(bd - ae)} \qquad \text{Eq. A3}$$

Finally $\lambda_2$ can be substituted back into Eq.12 to obtain $\lambda_1$.

$$\lambda_1 = \frac{2ce - 2bf}{(bd - ae)} \qquad \text{Eq. A4}$$

# Acknowledgements